\documentclass[rnote]{aa}
\usepackage[latin1]{inputenc}
\usepackage[T1]{fontenc}
\usepackage[francais]{babel}
\usepackage{graphicx}
\usepackage{amsmath,amsfonts,amssymb}
\usepackage{lmodern}
\usepackage{color}
\usepackage{natbib}

\begin{document}

\renewcommand{\refname}{References}
\renewcommand{\abstractname}{\large{Abstract}}

\title{Stability of rings around a triaxial primary}
   \author{Antoine Leh\'ebel
          \inst{1,2}
          \and
          Matthew S. Tiscareno\inst{2}
          }

   \institute{\'Ecole Normale Sup\'erieure de Cachan, 61 Avenue du Pr\'esident Wilson, 94230 Cachan, France    
         \and
                                Cornell University, Center for Radiophysics and Space Research, Ithaca, NY 14853, USA
             }
  \abstract
   {}
   {Generally, the oblateness of a planet or moon is what causes rings to settle into its equatorial plane. However, the recent suggestion that a ring system might exist (or have existed) about Rhea, a moon whose shape includes a strong prolate component pointed toward Saturn, raises the question of whether rings around a triaxial primary can be stable. We study the role of prolateness in the behavior of rings around Rhea and extend our results to similar problems such as possible rings around exoplanets.
}
   {Using a Hamiltonian approach, we point out that the dynamical behavior of ring particles is governed by three different time scales: the orbital period of the particles, the rotation period of the primary, and the precession period of the particles' orbital plane. In the case of Rhea, two of these are well separated from the third, allowing us to average the Hamiltonian twice. To study the case of slow rotation of the primary, we also carry out numerical simulations of a thin disk of particles undergoing secular effects and damping.}
   {In the case of Rhea, the averaging reduces the Hamiltonian to an oblate potential, under which rings would be stable only in the equatorial plane. This is not the case for Iapetus; rather, it is the lack of a prolate component to its shape that allows Iapetus to host rings. Plausible exoplanets should mostly be in the same regime as Rhea, though other outcomes are possible. The numerical simulations indicate that, even when the double averaging is irrelevant, rings settle in the equatorial plane on an approximately constant time scale.}
   {}

        \authorrunning{A. Leh\'ebel \and M.S. Tiscareno}
   \maketitle

\section{Introduction}

The oblateness of a planet (i.e., the slight flattening along its spin axis), characterized by the coefficient $J_{2}$, defines the Laplace plane in which potential rings may be stable \citep{TTN09}. However, some bodies, like moons orbiting their planet, are also elongated by tidal effects in the direction toward the object they are orbiting. This elongation is described by the coefficient $C_{22}$. Its effect on the stability of planetary rings is the subject of this work.

In 2008, the Cassini spacecraft observed some sharp symmetric drops in the flow of charged particles around Rhea \citep{Jones08}. This observation was interpreted as being due to the presence of narrow rings orbiting the moon. Furthermore, \citet{Schenk09} found a narrow equatorial color signature that was ascribed to the fallout of ring material and was cited as further evidence of Rhean rings.  However, \citet{Matt10} ruled out the current existence of rings with an in-depth imaging study of the equatorial region, after which \citet{Schenk11} suggested that the narrow equatorial color signature may still indicate the existence of rings in the past. 

Rhea would have been the first triaxial object known to possess a ring system. As for most moons close to their planet, both Rhea's $J_{2}$ and $C_{22}$ are significant. This work is motivated by the question of whether it is even possible for rings to be stable in such an environment.

The calculations we develop can be adapted to Iapetus, for which rings have been invoked to explain the origin of its equatorial mountain range \citep{Ip2006,Levison11,Dombard12}. Iapetus is different from Rhea in that it is simply oblate, but a comparison with our model remains interesting.

Some exoplanets, such as hot Jupiters, are likely to be drawn out by tidal effects in the direction of their star. This deformation, added to the fast rotation of the planet, must have complex effects on the presence of putative rings.

We will carry out a Hamiltonian analysis, following the model developed by \citet{DBS89} for Neptune polar rings. \citet{SC1984}  used a similar model to describe gas rings surrounding triaxial galaxies, which is better adapted to the current case of slowly rotating bodies.

In the next section, we show that the inertial frame is better adapted to the description of the motion than the rotating frame fixed on Rhea. In Sect. 3, we derive the perturbative Hamiltonian and show that the prolateness has no average effect. In Sect. 4, we investigate the case of Iapetus and exoplanets. Section 5 contains a computational simulation that allows us to study slower rotations.

\section{The choice of a frame}

We consider a test particle moving around Rhea on a nearly-Keplerian orbit perturbed by the shape of Rhea and the presence of Saturn. In this work, the particle's orbit is assumed to be circular. This is probably a good approximation for any putative rings.

\subsection{Definition of the frame and time scales}

\begin{figure}[ht]
\begin{center}
\includegraphics[width=6cm]{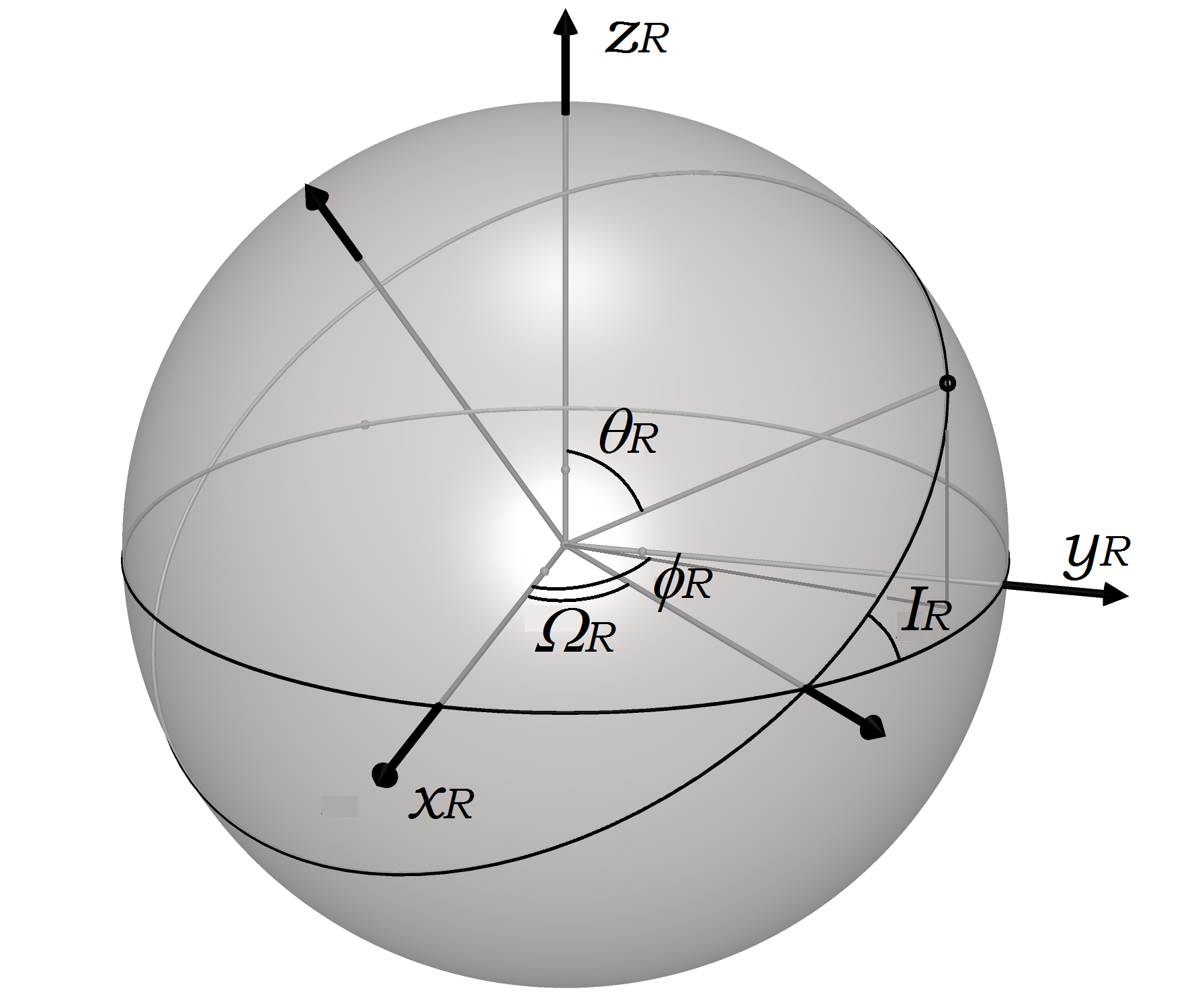}
\caption{Angles in the rotating frame}
\label{fig:defangl}
\end{center}
\end{figure}

We will first define a system of axes in which the expressions for the perturbations are simple (Fig. \ref{fig:defangl}). In general, $I$ is inclination, $\Omega$ is longitude of ascending node, and $\theta$ and $\phi$ are the particle's colatitude and longitude, respectively. The $x_\mathrm{R}$ axis always points toward Saturn during the motion of Rhea; therefore, it corresponds to Rhea's long axis and to the $C_{22}$ term. The $z_\mathrm{R}$ axis is taken parallel to Rhea's angular momentum, i.e., perpendicular to its orbital plane; it is the short axis. The $y_\mathrm{R}$ axis completes the other two  to form a direct basis. The angles in this frame will be labeled with a subscript R: $I_\mathrm{R}$, $\Omega_\mathrm{R}$, etc.

This frame is well adapted to express Rhea's gravitational potential, but it will not allow a simple description of the behavior of putative rings. In order to better understand the importance of the frame, we have to focus on the three relevant time scales of our problem: $\tau$, the period of one orbit of the test-particle around the moon; $\tau_\mathrm{R}$, the period of one orbit of the moon around the planet (identical to the rotation period of the moon); $\tau_\mathrm{p}$, the precession period of the test particle's orbit due to the non-sphericity of the moon. This precession is responsible for the settlement of the rings in preferred planes, so we can take this time as a low estimation of the settling time. 

If $\tau_\mathrm{R} \gg \tau$ and $\tau_\mathrm{R} \gg \tau_\mathrm{p}$, the particle has time to settle in the preferred planes relative to the moon, and the rotating frame coordinates is the most convenient. The rings turn with the moon during its motion.

If $\tau_\mathrm{p} \gg \tau_\mathrm{R}$, the rings remain steady in the inertial frame of Saturn during one period of Rhea. Therefore, on the time scale on which the particles can settle into rings, they only see an average potential: the rotating $C_{22}$ term averages to give an azimuthally symmetric shape to the satellite. In this case, the frame that rotates with Rhea is not the most convenient.

\subsection{The case of Rhea}

We estimate the different time scales as 

\begin{equation}
\left\{
\begin{array}{r c l}
\tau &=& \displaystyle\frac{2 \pi}{n} = \sqrt{\displaystyle\frac{4 \pi^2 a^3}{G M_\mathrm{R}}}\\
\\
\tau_\mathrm{R} &=& \displaystyle\frac{2 \pi}{n_\mathrm{R}} = \sqrt{\displaystyle\frac{4 \pi^2 a_\mathrm{R}^3}{G M_\mathrm{S}}}\\
\\
\tau_\mathrm{p} &=& \displaystyle\frac{2 \pi}{\dot{\Omega}} = \displaystyle\frac{2}{3 J_{2}} \sqrt{\displaystyle\frac{4 \pi^2 a^3}{G M_\mathrm{R}}} \left(\displaystyle\frac{a}{R_\mathrm{R}}\right)^2
\end{array}
\right.
\label{eq:constant}
,\end{equation}

\noindent where $a$ is the semi-major axis and $n = \sqrt{GM_\mathrm{R}/a^3}$ is the mean motion for the test particle on its orbit around Rhea, $a_\mathrm{R}$ and $n_\mathrm{R}$ are the same for Rhea on its orbit around Saturn, $M_\mathrm{R}$ and $M_\mathrm{S}$ are the respective masses of Rhea and Saturn, $R_\mathrm{R}$ is the radius and $J_2$ is the oblateness parameter for Rhea, and $G$ is Newton's constant. 

In $\tau_\mathrm{p}$, we estimated $\dot{\Omega}$ as $\dot{\Omega} = 3/2 n\: J_2\: (R_\mathrm{R}/a)^2$ from Equation 6.250 in \citet{MurDer}. This is valid when the only perturbation is oblateness, but it should give the correct precession rate to the first order in our situation, as the $J_2$ and $C_{22}$ are almost equally significant.

In the case of Rhea, $J_{2} = 889\;10^{-6}$, $M_\mathrm{R} = 2.308\;10^{21}$ kg, $R_\mathrm{R} = 764.3$ km, and $\tau_\mathrm{R} = 4.518\;d$ \citep{Anderson07}. If we take $a \simeq 1.5 \, R_\mathrm{R}$, as in \citet{Jones08}, the time scales are as given in Table \ref{table:time}. 

\begin{table}[ht]
\caption{Time scales of the problem for Rhea at $a = 1.5 \, R_\mathrm{R}$}
\begin{center}
\begin{tabular}{|l|c|}
\hline
$\tau$ & 0.227 d \\
\hline
$\tau_\mathrm{R}$ & 4.52 d \\
\hline
$\tau_\mathrm{p}$ & 384 d \\
\hline
\end{tabular}
\end{center}
\label{table:time}
\end{table}

Clearly, $\tau_\mathrm{p} \gg \tau_\mathrm{R}$. Thus, the most convenient frame is the inertial frame centered on Rhea that does not rotate. The transformation between the inertial and the rotational frames is just a rotation of angle $n_\mathrm{R} t$ around the $z$-axis. 

There is a factor of at least 50 between each time scale. This invites us to make a double averaging: the first  on the orbit of the particle around Rhea, and the second  on the orbit of Rhea around Saturn.

\section{The perturbative Hamiltonian}

\subsection{Derivation of the Hamiltonian}

The system is conservative, so the semi-major axis $a$ is constant. Therefore, an orbit is described by its inclination $I_\mathrm{R}$ and the position of the ascending node $\Omega_\mathrm{R}$. Another set of coordinates is relevant here, the particle's colatitude and longitude $(\theta_\mathrm{R},\phi_\mathrm{R})$, because the perturbative Hamiltonian is expressed in their terms. The time-evolution of $I_\mathrm{R}$ and $\Omega_\mathrm{R}$ is governed by \citep{Kaula1968} 

\begin{equation}
\left\{
\begin{array}{r c l}
\dot{I}_\mathrm{R} &=& \displaystyle\frac{1}{n a^2 \sin(I_\mathrm{R})} \frac{\partial H}{\partial \Omega_\mathrm{R}}\\
\\
\dot{\Omega}_\mathrm{R} &=& - \displaystyle\frac{1}{n a^2 \sin(I_\mathrm{R})} \frac{\partial H}{\partial I_\mathrm{R}}
\end{array}
\right.
\label{eq:evol}
.\end{equation}

All Hamiltonians will be written per unit of the test particle's mass. The most important perturbation for us is the one that results from the shape of Rhea \citep{SC1984}:

\begin{equation}
H_\mathrm{R} = n^2 R_\mathrm{R}^2 \left[J_{2} \left(\frac{3}{2} \cos^2 \theta_\mathrm{R} - \frac{1}{2}\right) - 3 C_{22} \sin^2 \theta_\mathrm{R} \cos 2 \phi_\mathrm{R}\right]
\label{eq:hr}
.\end{equation}

The particle also feels the effect of the gravitational field from Saturn. The quadrupole term (i.e., order 2 in the development in $a / a_\mathrm{R}$) is a good approximation for this perturbation. It is given by \citep{MurDer}

\begin{equation}
H_\mathrm{S} = \displaystyle\frac{n_\mathrm{R}^2 a^2}{2} (1 - 3 \sin^2 \theta_\mathrm{R} \cos^2 \phi_\mathrm{R}) - n_\mathrm{R}^2 a_\mathrm{R}^2
\label{eq:hsat}
.\end{equation}

We will now take the average of the perturbation over the duration $\tau = 2 \pi / n$ of one orbit of the particle around Rhea, $\overline{H} = (1/\tau) \int_0^\tau H \mathrm{d} t$, where $H = H_\mathrm{R} + H_\mathrm{S}$. Dropping the parts of the Hamiltonian that depend neither on $I$ nor on $\Omega$, we obtain 

\begin{multline}
\overline{H} = \displaystyle\frac{n^2 R_\mathrm{R}^2}{2}\left[ \displaystyle\frac{3 J_{2}}{2} \sin^2 I - 3 C_{22} \cos 2 (\Omega - n_\mathrm{R}t) \sin^2 I \right]\\
+ \displaystyle\frac{3 n_\mathrm{R}^2 a^2}{4} \sin^2 (\Omega-n_\mathrm{R} t) \sin^2 I.
\label{eq:HamR}
\end{multline}

We employ a similar method to average the Hamiltonian once more, over a time $\tau_\mathrm{R}$, which corresponds to one orbit of Rhea. Now the $C_{22}$ term vanishes, and the double-averaged Hamiltonian is

\begin{equation}
\overline{\overline{H}} = \displaystyle\frac{3}{4} \left(n^2 R_\mathrm{R}^2 J_{2} + \displaystyle\frac{n_\mathrm{R}^2 a^2}{2}\right) \sin^2 I
\label{eq:eff}
.\end{equation}

\subsection{Interpretation for Rhea}

The $\sin^2 I$ factor is characteristic of a $J_{2}$ perturbation. The presence of Saturn is equivalent to an increase in the oblateness of Rhea, and can be taken into account with an effective $J_{2}$. From Eq. \ref{eq:eff}, this effective $J_{2}$ is given by

\begin{equation}
J_{2}^{eff} = J_{2} + \displaystyle\frac{n_\mathrm{R}^2 a^2}{2 n^2 R_\mathrm{R}^2}
.\end{equation}

Close to the surface of Rhea, i.e., for $a = R_\mathrm{R}$, the second part is already equal to $375 \;10^{-6}$. This means that the effect of Saturn quickly becomes dominating when $a$ increases.

\section{Extension of the model}

\subsection{Validity of our approximations}

The averaging over several orbits of the particle around Rhea requires that $\tau_\mathrm{R}$ be  much larger than $\tau$. The other averaging, over several revolutions of Rhea necessitates that $\tau_\mathrm{p}$ be much larger than $\tau_\mathrm{R}$. We will study the most favorable possible case of a particle close to the moon's surface: $a \simeq R_\mathrm{R}$. With the definitions given in Eq. \ref{eq:constant}  we can rewrite the two previous conditions as

\begin{equation}
\left\{
\begin{array}{r c l}
\sqrt{q} &\ll& 1\\
\\
\displaystyle\frac{2 J_{2}}{3 \sqrt{q}} &\ll& 1
\end{array}
\right.
\label{eq:cond}
,\end{equation}

\noindent where $q$ is the rotation parameter defined by 

\begin{equation} 
q = \displaystyle\frac{n_\mathrm{R}^2}{n^2} = \displaystyle\frac{M_\mathrm{S}}{M_\mathrm{R}} \left(\displaystyle\frac{R_\mathrm{R}}{a_\mathrm{R}}\right)^3
.\end{equation}

\begin{figure}[h]
\begin{center}
\includegraphics[height=4.6cm]{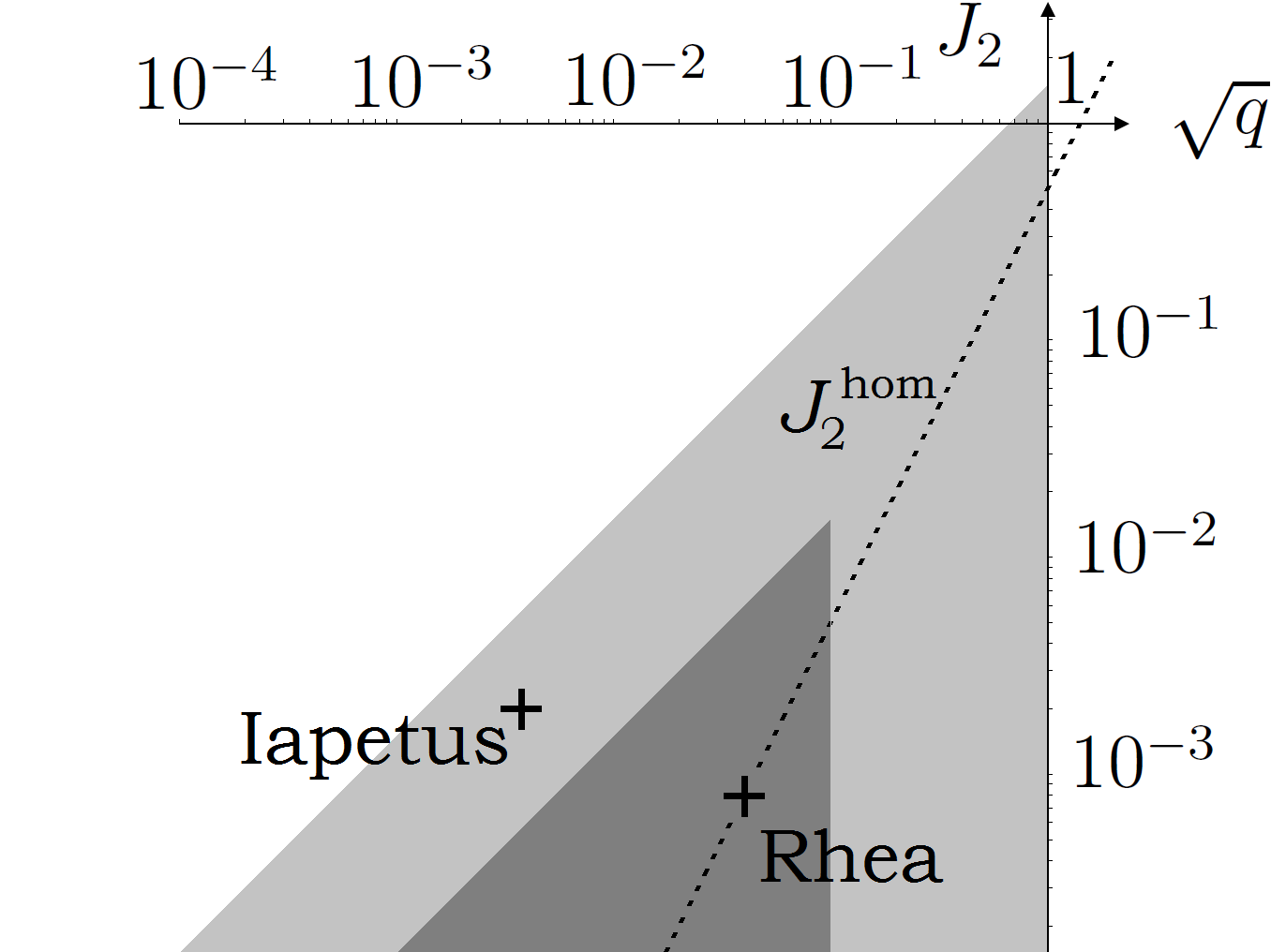}
\caption{Phase space in terms of Eqs. \ref{eq:cond}. Rhea lies in the dark gray zone, which means that we can safely apply our double-averaging method. Iapetus is not in that region. The dashed curve gives the position of homogenous, synchronous ellipsoidal objects.}
\label{fig:approx}
\end{center}
\end{figure}

Figure \ref{fig:approx} represents the space $(J_{2},\sqrt{q})$. The gray region corresponds to the region where Eqs. \ref{eq:cond} are satisfied, and the dark gray zone to the sets of parameters for which each pair of time scales is separated by at least 10x. For Rhea, $J_{2} = 889\;10^{-6}$ and $\sqrt{q}=2.7\;10^{-2}$ \citep{Anderson07}, so it lies in the dark region.

\subsection{Iapetus}

Iapetus' $J_2$ has never been measured, but \citet{Thomas07} give $a = 747.4$ km, $b = c = 712.4$ km, and mean radius $R_\mathrm{I} = 735.6$ km, yielding $J_2 = 1890\;10^{-6}$ \citep{Yoder95}, under the simplifying assumption that Iapetus is homogenous. With the definitions of the time scales given in Eq. \ref{eq:constant}, we calculate the values of Table \ref{table:timeiap} at $a =1.5\;R_\mathrm{I}$.

\begin{table}[ht]
\caption{Different time scales for Iapetus at $a = 1.5 \, R_\mathrm{I}$}
\begin{center}
\begin{tabular}{|l|c|}
\hline
$\tau$ & 0.248 d \\
\hline
$\tau_\mathrm{I}$ & 79.3 d \\
\hline
$\tau_\mathrm{p}$ & 197 d \\
\hline
\end{tabular}
\end{center}
\label{table:timeiap}
\end{table}

The condition required to make the averaging over several Iapetus rotations is not well satisfied: $\tau_\mathrm{I}/\tau_\mathrm{p} = 0.4$ (see Fig. \ref{fig:approx}). This would be an important effect if Iapetus did not have an azimuthal asymmetry. Thus, equatorial rings can be stable around Rhea and Iapetus, but for different reasons: for Iapetus it is the absence of a significant $C_{22}$ term, while for Rhea the $C_{22}$ term is neutralized by the fast revolution. If Iapetus' $C_{22}$ had the same magnitude as its $J_{2}$, polar rings would also be stable \citep{DBS89}, because the moon's rotation is slower than the precession of a particle in orbit around it.

\subsection{Exoplanets}

We now consider whether the non-spherical shape of some exoplanets (such as hot Jupiters) can destabilize a putative ring system.  In such an environment, any ring system would have to be composed of refractory material, not ice \citep{Chang11}. We choose a simple model of a homogenous synchronous planet at hydrostatic equilibrium. Therefore, we have \citep{MurDer}

\begin{equation}
J_{2} = \displaystyle\frac{q}{2}\\
\label{eq:hom}
.\end{equation}

This relation further simplifies the problem: the condition for averaging the exoplanet's shape over its rotation ($\tau_\mathrm{p} / \tau_\mathrm{R} \gg 1$) then becomes $\sqrt{q} \ll 3$, which is automatically satisfied with the first condition for averaging the particle along its orbit ($\tau_\mathrm{R} / \tau \gg 1$ or equivalently $\sqrt{q} \ll 1$). We represent the hydrostatic relation graphically in Fig. \ref{fig:approx}, with the dashed curve $J_{2}^\mathrm{hom} = q / 2$.

\begin{figure}[ht]
\begin{center}
\includegraphics[height=5.9cm]{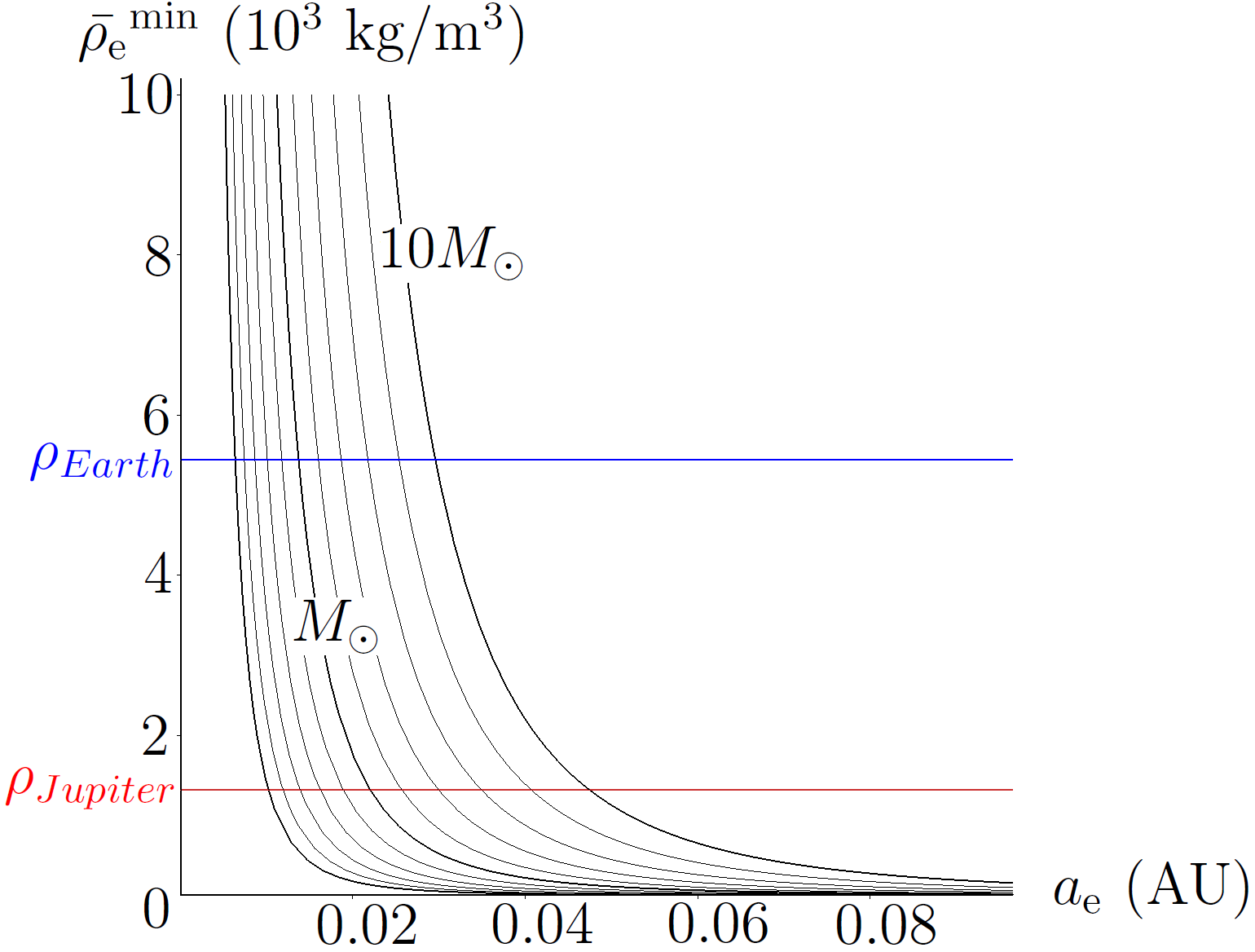}
\caption{The minimum density for which double-averaging is appropriate, as a function of the semi-major axis. Contours are for different masses of the central star.}
\label{fig:rholimit}
\end{center}
\end{figure}

We express $q$ in terms of more physical parameters: the mean density of the exoplanet $\overline{\rho_\mathrm{e}}$, the mass of the central star $M_{\ast}$, and the semi-major axis of the exoplanet $a_\mathrm{e}$. If we want a factor of 10 at least between each time scale, we can look for the limit parameter when the two others are fixed. The restrictive condition can then be written as 

\begin{equation}
\displaystyle\frac{M_{\ast}}{\overline{\rho_\mathrm{e}} a_\mathrm{e}^3} \lesssim \displaystyle\frac{\pi}{75}
.\end{equation}If $(M_{\ast},\! a_\mathrm{e})$ is set, there is a minimal mean density for the exoplanet below which we cannot apply our averagings: $\overline{\rho_\mathrm{e}}^{\: \mathrm{min}} \simeq 75 M_{\ast}/\pi a_\mathrm{e}^3$. Figure \ref{fig:rholimit} shows the minimum density of the exoplanet as a function of the semi-major axis for several given masses of the star from $0.1\;M_{\odot}$ to $10\;M_{\odot}$.

Figure \ref{fig:rholimit} indicates that above approximately 0.1 AU, all exoplanets are dense enough to satisfy the conditions. For $M_{\ast} = M_{\odot}$ and $a$ between 0.012 AU and 0.022 AU, only the planets with a density greater than Jupiter's are dense enough. For even smaller values of the semi-major axis, only telluric planets have the required density. Very few exoplanets orbit so close to their star that no reasonable density can fit the criteria. For hot Jupiters, the typical values are about $a_\mathrm{e} = 0.05$ AU and $M_{\ast} = M_{\odot}$, so the minimum density $\overline{\rho_\mathrm{e}}^{\: \mathrm{min}}$ equals 113~kg/m$^3$. This value is even lower than that of HD 209458 b, which is one of the least dense of the known exoplanets, with $\overline{\rho_\mathrm{e}}=350$~kg/m$^3$ \citep{Brown01}. For these exoplanets 
the conclusions about the role of the prolateness are the same as for Rhea.

\subsection{Hill's sphere limit}

In the case of exoplanets, the star also directly attracts the test particle. Hill's sphere radius can be estimated by $R_\mathrm{H} = a_\mathrm{e} (M_\mathrm{e} / 3 M_{\ast})^{1/3}$. It would have no meaning to consider rings outside Hill's sphere. Therefore, in a system that is able to shelter rings, Hill's radius must be greater than $a$, the semi-major axis of the orbiting particle. Setting $a=\alpha R_\mathrm{e}$, this condition can also be written in terms of $q$:  

\begin{equation}
q < \displaystyle\frac{1}{3 \alpha^3}
.\end{equation}

Even for very close rings, the maximum allowed value of $q$ is on the order of 0.1. This is the most unfavorable case, and yet it nearly already satisfies our hypotheses for the double averaging ($\sqrt{q} \ll 1$). To sum up, we should be careful when $q$ increases too much, but in these cases the attraction of the central body is likely to prevent any rings from forming.

\section{Computational simulation for slower rotations}

\subsection{Model and results}

Not all planets are homogenous or at hydrostatic equilibrium. Iapetus is a typical example of a planet whose $J_{2}$ is too large to be caused by rotation at its distance to Saturn. \citet{Thomas07} called this a ``fossil bulge'', based on the likely idea that it is left over from a time when Iapetus rotated faster. Therefore we consider what happens when the rotational period of a moon (e.g., Rhea) $\tau_\mathrm{R}$ is not short compared with the precession time $\tau_\mathrm{p}$. The Hamiltonian to consider is then the one given by Eq. \ref{eq:HamR}. Given its complexity, we will use a simulation.

Our simulation will describe the behavior of a thin disk of interacting particles. To take into account these interactions, we will simply give our particle disk a kinematic viscosity $\nu$, as in \citet{DBS89} and \citet{SC88}. We  modify the system of Eqs. \ref{eq:evol} by adding a dissipative term $(\dot{I}_\mathrm{D},\dot{\Omega}_\mathrm{D})$ \citep[see Eqs. 9 and 10 in][]{DBS89}:

\begin{equation}
\left\{
\begin{array}{r c l}
\dot{I}_\mathrm{D} &=& \displaystyle\frac{\nu}{2} \left[\displaystyle\frac{\partial^2 I}{\partial a^2} - \sin I \cos I \left(\displaystyle\frac{\partial \Omega}{\partial a} \right)^2 \right]\\
\\
\dot{\Omega}_\mathrm{D} &=&   \displaystyle\frac{\nu}{2} \left[\displaystyle\frac{\partial^2 \Omega}{\partial a^2} + 2 \cot I \displaystyle\frac{\partial \Omega}{\partial a} \displaystyle\frac{\partial I}{\partial a} \right]
\end{array}
\right.
.\end{equation}

We assume that $a$ is constant. Indeed, the settling times are much shorter than the amount of time where the inflow is significant \citep{DBS89}. The ring is then modeled as a set of 100 annular elements. Forty kilometers separate the innermost annulus from the outermost, and all annuli have the same initial inclination. Then, we take a time step of $\tau_\mathrm{R} / 100$. The kinematic viscosity is estimated to be about 1~m$^2$/s, using the model of \citet{SC88}, with ice particles of millimeter size. The settling time is denoted $\tau_\mathrm{s}$.

Testing our simulation with an object having the characteristics of Rhea, we find satisfactory results. No matter what the initial inclination is, the rings always settle into the equatorial plane, as expected. The settling time we admitted for our theoretical study was equal to the precession period $\tau_\mathrm{p}$. The simulation gives settling times $\tau_\mathrm{s}$ which are closer to ten times $\tau_\mathrm{p}$. However, our goal was only to obtain a theoretical behavior for the rings, not to predict the detailed settling process. Such a difference is therefore acceptable.

We can now try to vary the parameters of the problem, and especially to study what happens when $\tau_\mathrm{R} \simeq \tau_\mathrm{p}$. We will decrease the rotational frequency $n_\mathrm{R}$. This is equivalent to moving Rhea outward from Saturn while retaining its current shape ($J_2$ and $C_{22}$) as a fossil bulge. We will call this new frequency $n_\mathrm{R}^\mathrm{var}$ to avoid confusion with Rhea's proper characteristics.

We find that the behavior of the rings is roughly independent of $n_\mathrm{R}^\mathrm{var}$. The rings still settle in the equatorial plane during a nearly constant duration $\tau_\mathrm{s}$. Table \ref{table:taus} shows that $\tau_\mathrm{s}$ is not dependent on the rotational speed.

\begin{table}[ht]
\caption{Settling time at several rotational frequencies.}
\begin{center}
\begin{tabular}{|l|l|}
\hline
\rule[.2cm]{0cm}{.2cm}
$\tau_\mathrm{s}$ ($10^3$ days) & $n_\mathrm{R}^\mathrm{var}/n_\mathrm{R}$ \\
\hline
\rule[.16cm]{0cm}{.16cm}
$5.6$ & $10^{-2}$ \\
\hline
\rule[.16cm]{0cm}{.16cm}
$5.3$ & $10^{-1}$ \\
\hline
\rule[.16cm]{0cm}{.16cm}
$5.3$ & $1$ \\
\hline
\rule[.16cm]{0cm}{.16cm}
$5.4$ & $10$ \\
\hline
\rule[.16cm]{0cm}{.16cm}
$5.0$ & $10^{2}$ \\
\hline
\end{tabular}
\end{center}
\label{table:taus}
\end{table}

\subsection{Caveat}

The simulation results should be interpreted with some caution. While the simulation robustly reproduces secular effects, it may not represent any effects that are not azimuthally averaged, such as any imparting of eccentricity to the ring or any resonant effects due to the rotating character of the $C_{22}$ term.

\section{Discussion}

The separation between the different time scales in the case of Rhea allows us to carry out a double averaging and obtain a more understandable Hamiltonian. We have shown that rings can settle only in the equatorial plane. This rules out the possibility of rings in a region around Rhea that would not have been carefully investigated. The results of \citet{Matt10} then seem to definitively exclude the present existence of these rings.

Our model turned out to be interesting for other celestial objects. The most likely objects to have a shape close to Rhea's are exoplanets orbiting close to their stars. Nearly all of them should be suitable for our model, except for those orbiting at less than 0.02 AU from their star. A weak density could put such rare planets into a more complex regime. On the other hand, such close exoplanets are unlikely to shelter rings owing to their small Hill spheres. Therefore, results obtained for Rhea can  be extended to almost all synchronous exoplanets. 

Finally, we carried out a computational simulation focused on what happens when the rotational period becomes as significant as the precession time. This happens when the $J_2$ is larger than it should be, as it is in the case of Iapetus. Though it neglects all effects that are not azimuthally averaged, our simulation indicates that, over a wide range of rotational periods, a thin particle disk will become rings in the equatorial plane. The duration for the settling is roughly independent of the rotational period.

Thus, in all cases that were studied, the prolate shape seems to have no determining effect on the shape of the rings, keeping in mind that  classical shape models for the exoplanets were assumed. Nothing prevents the discovery of more original shapes, in particular higher prolate deformations allowing more imaginative ring systems.\\

We thank Joe Burns, Phil Nicholson, Matt Hedman, and Tom Steiman-Cameron for helpful conversations.  We acknowledge funding from the \'Ecole Normale Sup\'erieure de Cachan and from NASA Outer Planets Research (NNX10AP94G).

\bibliographystyle{apalike}
\bibliography{rings_triaxial_print_4}

\end{document}